\documentclass[aps,prl,twocolumn,showpacs,a4paper,unsortedaddress,amsmath,amssymb,byrevtex]{revtex4}

\usepackage[latin1]{inputenc}
\usepackage{graphicx}
\usepackage{dcolumn}
\usepackage{bm}
\usepackage{hyperref}
\usepackage{times}

\begin{document}
\preprint{ffuov/02-01}

\title{Conductance oscillations in zigzag platinum chains}

\author{V. M. Garc\'{\i}a-Su\'arez$^{1,3}$}
\author{A. R. Rocha$^2$}
\author{S. W. Bailey$^3$}
\author{C. J. Lambert$^3$}
\author{S. Sanvito$^2$}
\author{J. Ferrer$^1$}
\affiliation{$^1$ Departamento de F\'{\i}sica, Universidad de Oviedo, 33007 Oviedo, Spain}
\affiliation{$^2$ Physics Department, Trinity College, Dublin 2, Ireland}
\affiliation{$^3$ Department of Physics, Lancaster University, Lancaster, LA1 4YB, U. K.}

\date{\today}

\begin{abstract}
Using first principles simulations we perform a detailed study of
the structural, electronic and transport properties of monoatomic
platinum chains, sandwiched between platinum electrodes. First, we
demonstrate that the most stable atomic configuration corresponds
to a zigzag arrangement that gradually straightens as the chains
are stretched. Secondly, we find that the averaged conductance
shows slight parity oscillations with the number $n$ of atoms in the
chain. Additionally, the conductance of chains of fixed $n$
oscillates as the end atoms are pulled apart, due to the gradual
closing and opening of conductance channels as the chain
straightens.
\end{abstract}


\maketitle

The existence of single atom chains was demonstrated some time ago
using the scanning tunneling microscope (STM) and
mechanically-controllable break junctions (MCBJ)
\cite{Ohn98} where a quantized conductance close to
$G_0=2\,e^2/h$ for gold was measured, in agreement with previous
theoretical predictions \cite{Fer88}. Since then, a number of
experiments \cite{Smi01,Smi03} and theoretical calculations
\cite{Bah01} have proved that the 5d elements Au, Ir and Pt can be
used to produce monoatomic chains. 
For gold chains
\cite{Ohn98}, the average distance between conductance peaks
was found to be 2.5 \AA\ with the conductance $G$ of the last
plateau being very close to $G_0$. The ensemble-averaged
conductance also shows small oscillations around $G_0$ as the
length of the chain increases \cite{Smi03}. 

For platinum, the average distance between the peaks in the
length-histograms is about 1.9-2.3 \AA~ \cite{Smi01,Smi03}. 
In contrast with gold, the conductance is no longer an
integer multiple of $G_0$, but instead decreases from 1.6 $G_0$ to
1.2 $G_0$ as the length of the chain increases \cite{Smi03}. In
addition, significant conductance oscillations are superimposed on
top of this decreasing trend, which have been attributed to a
parity effect.
\cite{Smi03,Nie03}.
The rich behavior of platinum compared with gold arises from the
larger number of conduction channels at the Fermi energy ($E_F$)
due to the presence of d bands\cite{vega}.

To understand the structural, electronic and transport properties
of platinum chains, we have performed a complete series of first
principles simulations of platinum chains attached to platinum fcc
leads. We have employed our newly developed code
SMEAGOL\cite{Roc04,Natmat}, which calculates the density matrix
and the transmission coefficients of a two probe device using the
non-equilibrium Green's Function formalism (NEGF) \cite{Kel65}.
The scattering potential is  calculated self-consistently using
the SIESTA implementation of density functional theory
\cite{Sha65,Sol02,SIE}. We have approximated the exchange and
correlation potential by the Local Density Approximation (LDA),
since we have found that provides a slightly more accurate description of the
structural and conducting properties of bulk and infinite
platinum chains than GGA \cite{Scuseria}. We have not included in our 
calculations spin polarization which is only relevant for very stretched
chains \cite{Delin}. Our main result is that the most stable
arrangement of platinum chains corresponds to zigzag
configurations, that are straightened as the chains are stretched.
The conductance decreases from 1.6 $G_0$ to 1.2 $G_0$ upon chain
stretching, showing slight parity oscillations. Aditionally, we
also find geometry-induced oscillations as the electrodes are
pulled apart. Such oscillations are due to the gradual opening and
subsequent closing of transport channels as chains with a fixed
number of atoms $n$ are straightened.

\begin{figure}
\includegraphics[width=8cm,height=7cm]{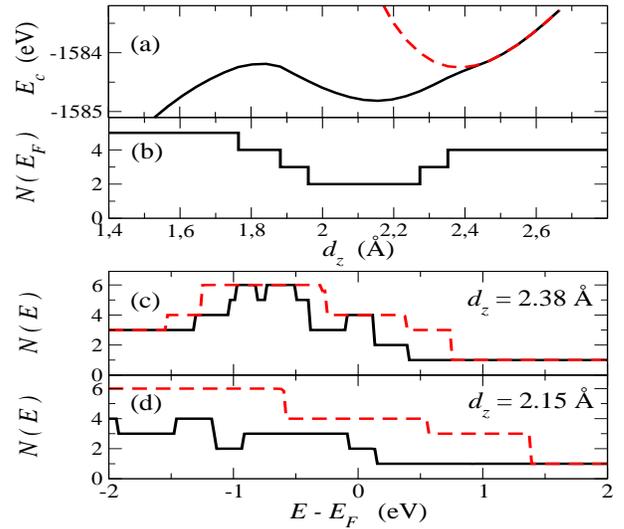}
\caption{\label{Fig1} (color online) (a) Cohesive energy $E_c$ and
(b) number of open scattering channels at the Fermi energy $N(E_F)$
of platinum chains as a function of $d_z$. Number of open scattering
channels $N(E)$ as a function of energy $E$ calculated
at (c) $d_z=2.38$ and (d) $d_z=2.15$ \AA\ for zigzag (solid line) and
linear (dashed line) chains.}
\end{figure}

We start by discussing the case of infinite platinum chains, which
is useful to understand the basic physics of the constriction when
a long chain is formed. We use a two atom unit cell with periodic
boundary conditions along the three spatial directions. The cell is
large enough along the $x$ and $y$ axes to avoid spurious interactions
between the chain in the central unit cell and its images. We simulate two
kinds of chains: linear chains, where the atoms are constrained to
lie along the $z$ axis, and zigzag chains, where forces are
allowed to relax along the three spatial coordinates. 
We find
that zigzag chains are more stable than linear chains, as it can
be seen from the cohesive energy of Fig. \ref{Fig1}(a). The
equilibrium distances along the $z$ axis are $d_{z,\tt eq}$ = 2.15
and 2.38 \AA\ for the zigzag and linear chains, respectively. In
the zigzag arrangement the Pt-Pt bonds are located in the $xz$
plane and make a 24.8 degrees angle with the $z$ axis. This means
that the interatomic distance for the zigzag configuration is
$d=2.37$ \AA, which is almost the same as for the linear case.
These angles decrease almost linearly as the chains are stretched,
and become approximately zero for $d_z$ larger than 2.5 \AA.
Interestingly, if the distance is reduced below the zigzag minimum
the system falls into another stable configuration with a ladder
arrangement, similar to that predicted by Sen et al. \cite{Sen01}.

Fig. \ref{Fig1}(b) shows the number of open scattering channels
$N(E_F)$ at the Fermi energy $E_F$ of an infinite zigzag chain, as
a function of $d_z$. The figure reveals that $N(E_F)=5$ for small
$d_z$, but decreases to $N(E_F)=2$ as the chain is stretched
beyond 1.95 \AA, and maintains that value for quite a large range
of distances. For $d_z$ somewhat larger than 2.30 \AA, $N(E_F)$
increases in two steps to 4 $G_0$, and stays constant thereafter
until the chain breaks. To understand which channels are opening
and closing, we have studied the projected density of states
(PDOS) for chains at different stretching. As limiting cases we
compared a linear chain at $d_z$ = 2.38 with a zigzag chain at
$d_z$ = 2.15 \AA. For the linear chain we find that the d$_{xy}$
and d$_{x^2-y^2}$ orbitals are completely filled, while the
hybridized s, d$_{xz}$, d$_{yz}$ and  d$_{z^2-r^2}$ orbitals have
all finite weight at $E_F$, leading to four open channels, as
shown in Fig. \ref{Fig1}(c). Atoms in zigzag chains at $d_z$ =
2.38 \AA$\,$ make small angles with the $z$-axis at this distance
and therefore, the $d_z$-dependence of the PDOS and $N(E_F)$ look
both fairly similar to those of the linear chain.
In contrast, the number of open channels $N(E)$ as a function of
energy $E$, shown in Fig. \ref{Fig1}(d), and the PDOS of zigzag
chains at $d_z$ = 2.15 \AA\ look rather different from those of
the linear chain. The zigzag chain has only two, and not four,
open channels at the Fermi energy, corresponding to a mixture of
all d-orbitals, since now the s-orbital is completely filled,
while the d$_{xy}$ and d$_{x^2-y^2}$ have moved up in energy. This
analysis of infinite chains can explain why in some experiments
the conductance increases when the electrodes separate, because
there is a gradual transition from a zigzag to a linear
configuration upon the stretching of the chain, with the linear
chain presenting larger number of open scattering channels.

\begin{figure}
\includegraphics[width=6cm,height=4cm]{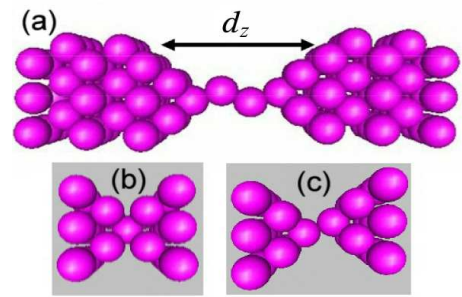}
\caption{\label{Fig2} (color online) The different chains
connecting (001) oriented fcc leads studied in this paper:
(a) Four-atoms chain, (b) single atom contact and (c) 2-atom chain.
$d_z$ is the distance between leads.}
\end{figure}

\begin{figure}
\includegraphics[width=7cm,height=4cm]{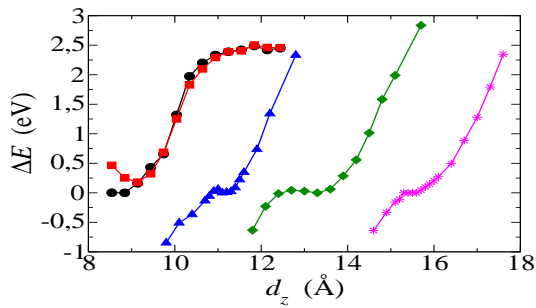}
\caption{\label{Fig3} (color online) Cohesion curves of the chain
plus leads systems, for zigzag and linear chains of two atoms
(circles and squares, respectively), and zigzag chains containing
3, 4 and 5 atoms (triangles, diamonds and stars, respectively). All
curves have been shifted in energy in order to align the local minima.
}
\end{figure}

This simplified picture cannot explain however the features
that are present in a real experiment, where the contact to the
electrodes and rearrangement of the leads near the surface can
decrease the transmission of channels or even close them. To
investigate this possibility, we have simulated finite chains of
various lengths (between 1 and 5 atoms), attached to fcc platinum
leads oriented along the (001) direction (see Fig.~\ref{Fig2}).
The leads are composed of repeated slices of $3\times 3$ atoms and
are connected to the chain through a square of 4 atoms, i.e.
through the fcc (001) hollow site. In order to get rid of
undesirable oscillations in the transmission coefficients, we have
used periodic boundary conditions along the $xy$ plane and summed
over 12 $k$-points. To treat the contact region self-consistently,
three atomic planes of bulk platinum are included in each side
of the scattering region\cite{planes}.
To calculate the most stable configuration for each chain, we
perform the relaxation by keeping fixed the bulk Pt leads and
relaxing the apexes of the point contact. In this way the
relaxation is performed over the chains and the two $2\times 2$
planes forming the hollow site. The ground state energy is therefore
calculated as a function of the distance $d_z$ between the outer
slices (unrelaxed) as indicated in figure \ref{Fig2}. In
Fig.~\ref{Fig3} we plot the cohesion curves for zigzag atomic
chains with a number of atoms ranging between 2 and 5.

The ground state energy of the single atom contact (not shown) as a
function of distance is a parabola whose minimum is located at the
equilibrium distance $d_{z,\tt eq}=6.3$ \AA. A parabolic
dependence is also found when we place two atoms facing each
other, with $d_{z,\tt eq}=9.1$ \AA. However, if we allow the outer
planes to move along the $x$ and $y$ directions, these two atoms
achieve a zigzag configuration that has a lower energy, as shown
in Fig.~\ref{Fig3}. This zigzag arrangement is actually a local
minimum, as in the case of chains of infinite length. For short
distances, the atoms at the apex gain energy by forming further
chemical bonds to other atoms at the electrodes. The same kind of
curves are observed for 3, 4 and 5 atoms, with $d_{z,\tt eq}=$
11.4, 13.3 and 15.4 \AA, respectively. We find that the difference
in lengths between chains of $n$ and $n+1$ atoms lies in the range
1.9 - 2.1 \AA, in very good agreement with experiments
\cite{Smi03}. The only configuration which does not fit into this
pattern is the single atom contact, whose equilibrium distance is
2.8 \AA~ away from the two-atom chain. Indeed, the single atom
contact does not constitute a chain by itself, since a large
number of bonds link the central atom to its neighbors, which must
be broken simultaneously in order to snap the chain. In contrast,
atomic chains can be topologically characterized as broken by
cutting just one single bond, which leads  us to ascribe the peaks
found in length histograms \cite{Smi03} to chains 2-, 3-, 4- and
5-atoms long. Fig. \ref{Fig3} also indicates that the  region of
stability of the $(n+1)-$atom chain begins at a distance where the
the $n-$atom chain is very stretched and therefore close to being
broken. This fact helps understand why the  first two peaks found
in length-histograms, that correspond to chains with two or three
atoms, are much higher than those attributed to chains of four or
five atoms \cite{Smi01,Smi03}, because there is a significant
probability that an $n$-atom chain will break, before the
$n+1$-atom chain forms. At a more detailed level, we also find
that the angles between the atomic bonds and the chain axis are
small for short chains (9.1 degrees for the 3-atom chain), but
increase with the chain length (21.6 degrees for 4-atom chain).
The case of the 5-atom chain is more complicated: the two atoms
joining the leads make angles of 36.3 degrees, while those in the
middle have angles equal to 16.1 degrees.

\begin{figure}
\includegraphics[width=7cm,height=4cm]{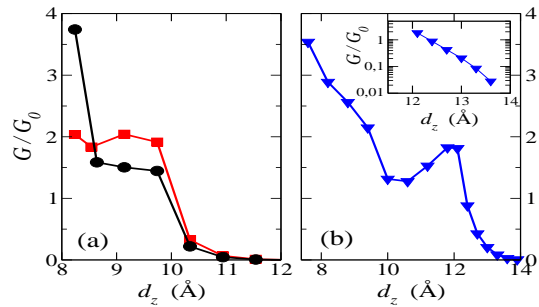}
\caption{\label{Fig4} (color online) Conductance of platinum
chains. (a) 2 atoms in zigzag or linear arrangement (circles or squares,
respectively). (b) 3 atoms; in the inset is plotted the tail of
the curve in a logarithmic scale, that shows the tunneling
behavior when the chain breaks.}
\end{figure}

Moving to transport, we observe that the single atom contact
behaves very differently from any other, with a conductance of
about 5.9 $G_0$ at the equilibrium distance, which decreases
almost linearly to $G=3.5$ $G_0$ until the contact breaks. These
values are very different from those of reference \cite{Nie03},
where the pyramid-like structure of the contact was not included,
probably leading to an overestimate of the interatomic distance. For the
2-atom chain we find a conductance of 1.5 $G_0$ at the equilibrium
distance in the zigzag arrangement and a conductance ranging
between 1.8 and 2.0 $G_0$ in the linear case, as shown in Fig.
\ref{Fig4}(a). The second value can be associated with the return
conductance measured when a contact is made again after the system
breaks \cite{Nie03}. We therefore propose that the first
configuration established between two atoms should be linear when
both tips are brought together.

For larger chains, we obtain a non-monotonic behavior of the
conductance as a function of $d_z$. When the separation between
the leads increases from a compressed configuration, the
conductance initially decreases and then exhibits a plateau at
around the equilibrium distance. Under further expansion it grows
again (for stretching of about 1 \AA\ beyond the equilibrium
separation) and finally decreases exponentially. An example is
shown in Fig. \ref{Fig4} (b), for $n=3$. This behavior can be
understood as follows. For small distances, atoms in the chain are
very close to each other and there is a large number of open
channels. As the distance increases, the transmission through many
of these channels is reduced and the conductance decreases until a
value between 1.0 and 1.5 $G_0$ is reached, where the chain has a
zigzag configuration. If the chain is stretched further, the
conductance increases again, following the evolution from a zigzag
to a linear chain, until a small plateau with a conductance of 2
$G_0$ is formed. Finally, the chain enters the exponential
tunneling regime, as shown in the inset to Fig. \ref{Fig4} (b),
whose onset signals the point where the chain would break or a new
atom would enter the chain.

\begin{figure}
\includegraphics[width=7cm,height=6cm]{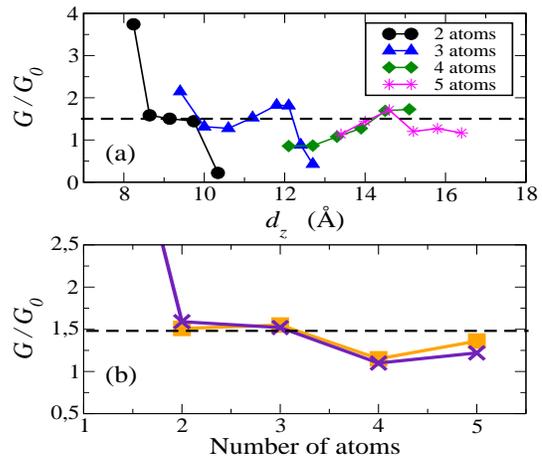}
\caption{\label{Fig5} (color online) (a) Evolution of the conductance
of 2-, 3-, 4- and 5-atoms chains as a function of $d_z$ (circles,
triangles, diamonds and stars, respectively). (b) Conductance
vs. the number of atoms in the chain measured at the equilibrium
distance (crosses) and averaged over a range of two Angstrom about
such distance (squares).
}
\end{figure}

To try to make contact with the experiments of Ref. \cite{Smi03},
the conductances of all chains are gathered together in Fig. 5 (a).
We find clear oscillations with a periodicity equal to the interatomic
distance, which have a structural origin. These oscillations are due to
the gradual closing and opening of channels, which occurs as the angles between the
atomic bonds and the $z$-axis increase and decrease. For a
stretched chain, the angles increase if a new atom enters the
chain, and subsequently decrease as the chain is further
stretched. Moreover the size of the conductance is linked to the
biggest angle subtended by atoms in the middle of the chain.
Larger chains tend to have bigger angles and lower conductances.
One exception is the 5-atom chain, whose conductance, 1.2 $G_0$,
is slightly bigger than that of the four-atom chain (1.0 $G_0$).
This is also easily explained by our calculations, since atoms in
the middle of the 5-atoms chains have a smaller angle than those
of four atoms chains (16.1$^0$ versus 21.6$^0$).

The large geometry oscillations of Fig. 5(a) may mask those due to the parity
effect, which have a larger periodicity but a smaller amplitude.
The experimental conductance curves shown in Ref. \cite{Smi03}
are obtained by averaging individual conductance traces over thousands of
pulling circles. We therefore expect that this averaging should smooth out short-distance fluctuations, such as the large geometry oscillations discussed
above, and uncover parity oscillations, that have a longer wavelength. To explore
such a possibility, we have plotted in Fig. 5 (b) the conductance of each chain,
taken at its equilibrium distance. We find that the large geometry oscillations
have been completely washed out, leaving only a small oscillation that has a
periodicity equal to twice the interatomic distance. To further check the stability
of these results, we have also plotted in Fig. 5 (b) the conductance of each
chain, but now averaged over a range of lengths of about two Angstroms around
each equilibrium distance. We believe that such a procedure may reproduce the
most important features of the experimental averaging of conductance curves. Interestingly, both curves almost
overlap.  The remaining curve actually shows two peaks, corresponding
to chains with three and five atoms. We note that the position of the conductance
peaks, the overall shape of the curve and the magnitude of the conductance are in
excellent quantitative agreement with the results of Ref. \cite{Smi03}.
In view of the above results, we propose that the geometry oscillations
seen in our simulations are averaged over in the experiments by
Smit et al. \cite{Smi03}, leaving only the smaller-amplitude, but robust
parity oscillations.

In summary, we have shown that monoatomic platinum chains have a
zigzag structure, both in the case of perfect infinite chains,
where the conductance is halved compared to the linear
configuration, and the case of chains between two fcc (001)
electrodes. We find that atomic chains are composed of at least two
atoms, while the single atom contact shows significantly different
structural and transport features. Our calculations explain
the negative slope in the conductance that is found for larger chains,
and accurately reproduce the oscillations given by parity effects
found in the experiments of Ref. \cite{Smi03}.

\begin{acknowledgments}
JF wishes to acknowledge conversations with J. J. Palacios and L.
Fern\'andez-Seivane. 
We acknowledge the finantial support provided by the
European Comission, the spanish Ministerio
de Eduaci\'on y Ciencia, the british EPSRC, Royal Society and NWDA, and
the irish Enterprise Ireland.
\end{acknowledgments}


\begin{thebibliography}{99}
\bibitem{Ohn98} H. Ohnishi, Y. Kondo, and K. Takayanagi, Nature (London)
{\bf 395}, 780 (1998). A. I. Yanson, G. Rubio Bollinger, H. E. van der
Brom, N. Agrait, and J. M. van Ruitenbeek, Nature (London) {\bf 395},
783 (1998).
\bibitem{Fer88} J. Ferrer, A. Mart\'{\i}n-Rodero, and F. Flores,
Phys. Rev. B {\bf 38}, R10113 (1988).
\bibitem{Smi01} R. H. M. Smit, C. Untiedt, A. I. Yanson, and J. M. van
Ruitenbeek, Phys. Rev. Lett. {\bf 87}, 266102 (2001).
\bibitem{Smi03} R. H. M. Smit, C. Untiedt, G. Rubio-Bollinger, R. C. Segers,
and J. M. van Ruitenbeek, Phys. Rev. Lett. {\bf 91}, 076805
(2003).
\bibitem{Bah01} S. R. Bahn and K. W. Jacobsen, Phys. Rev. Lett.
{\bf 87}, 266101 (2001).
\bibitem{Nie03} S. K. Nielsen, Y. Noat, M. Brandbyge, R. H. M.
Smit, K. Hansen, L. Y. Chen, A. I. Yanson, F. Besenbacher, and J.
M. van Ruitenbeek, Phys. Rev. B {\bf 67}, 245411 (2003).
\bibitem{vega} L. de la Vega, A. Mart\'{\i}n-Rodero, A. Levy Yeyati and 
A. Sa\'ul, Phys. Rev. B {\bf 70} 113107 (2004).
\bibitem{Roc04} A. R. Rocha, V. M. Garc\'{\i}a-Su\'arez, S. W. Bailey, C.
J. Lambert, J. Ferrer, and S. Sanvito, condmat/0510083.
\bibitem{Natmat}A.~R.~Rocha, V.~M.~Garc\'{\i}a-Su\'arez, S.~W.~Bailey, C.~J.~Lambert,
J.~Ferrer and S.~Sanvito, Nature Materials {\bf 4}, 335 (2005).
\bibitem{Kel65} L. V. Keldysh, Sov. Phys. JETP {\bf 20}, 1018 (1965).
\bibitem{Sha65} W. Kohn and L. J. Sham, Phys. Rev. {\bf 140}, A1133 (1965).
\bibitem{Sol02} J. M. Soler, E. Artacho, J. D. Gale, A. Garc\'{\i}a, J. Junquera,
P. Ordej\'on, and D. S\'anchez-Portal, J. Phys.: Condens. Matter
{\bf 14}, 2745 (2002).
\bibitem{SIE} We have used a double-$\zeta$ polarized basis set (DZP), where the cutoff
radii were optimized variationally.
\bibitem{Scuseria} J. Uddin and G. Scuseria, Phsy. Rev. B {\bf 72} 035101 (2005).
\bibitem{Delin} A. Delin and E. Tosatti, Phys. Rev. B {\bf 68}, 144434 (2003); J.
Fern\'andez-Rossier, D. Jacob, C. Untiedt and J. J. Palacios, condmat/0510153.
\bibitem{Sen01} P. Sen, S. Ciraci, A. Buldum, and I. P. Batra,
Phys. Rev. B {\bf 64}, 195420 (2001).
\bibitem{planes} We have performed several additional simulations including
either four and five atomic planes in the scattering region. These show
almost identical elastic properties and conductance.
\end{thebibliography}
\end{document}